\documentclass[aps,prb,twocolumn,superscriptaddress]{revtex4-1}
\usepackage{graphicx}
\usepackage{amssymb}
\usepackage{amsmath}
\usepackage{hyperref}
\usepackage{color}

\newcommand{\pd}{\partial}

\newcommand{\mr}[1]{\mathrm{#1}}

\begin{document}

\title{Persistent Corner Spin Mode at the Quantum Critical Point of a Plaquette Heisenberg Model}

\author{Yining Xu}
\affiliation{College of Physics and Electronic Engineering, Chongqing Normal University, Chongqing 401331, China}
\author{Chen Peng}
\affiliation{Kavli Institute for Theoretical Sciences and CAS Center for Excellence in Topological Quantum Computation, University of Chinese Academy of Sciences, Beijing 100190, China}
\author{Zijian Xiong}
\email{xiongzjsysu@hotmail.com}
\affiliation{Department of Physics, and Center of Quantum Materials and Devices, Chongqing University, Chongqing, 401331, China}
\affiliation{Chongqing Key Laboratory for Strongly Coupled Physics, Chongqing, 401331, China}
\author{Long Zhang}
\email{longzhang@ucas.ac.cn}
\affiliation{Kavli Institute for Theoretical Sciences and CAS Center for Excellence in Topological Quantum Computation, University of Chinese Academy of Sciences, Beijing 100190, China}

\begin{abstract}
Gapless edge states are the hallmark of a large class of topological states of matter. Recently, intensive research has been devoted to understanding the physical properties of the edge states at the quantum phase transitions of the bulk topological states. A higher-order symmetry-protected topological state is realized in a plaquette Heisenberg model on the square lattice. In its disordered phase, the lattice with an open boundary hosts either dangling corner states with spin-$1/2$ degeneracy characterizing the topological phase, or nondangling corner states without degeneracy, which depends on the bond configuration near the corners. In this work, we study the critical behavior of these corner states at the quantum critical point (QCP), and find that the spin-$1/2$ corner state induces a new universality class of the corner critical behavior, which is distinct from the ordinary transition of the nondangling corners. In particular, we find that the dangling spin-$1/2$ corner state persists at the QCP despite its coupling to the critical spin fluctuations in the bulk. This shows the robustness of the corner state of the higher-order topological state.
\end{abstract}

\maketitle

\section{Introduction}

Phase transitions and critical phenomena are among the central topics of condensed matter and statistical physics. The canonical understanding of continuous phase transitions, the Landau-Ginzburg-Wilson (LGW) theory \cite{Cardy1996scaling}, has been developed since the 1940s based on the notions of spontaneous symmetry breaking and the renormalization group (RG) transformations. Most of classical critical points and a large part of quantum critical points (QCPs) belong to the Wilson-Fisher universality classes, which only depend on the broken symmetry at the phase transition and the space (or spacetime) dimensionality. Quantum phase transitions beyond the LGW paradigm, such as the deconfined quantum critical point \cite{Senthil2004b, Senthil2004a}, are under intensive research, in which the quantum Berry phases and topological terms play a prominent role \cite{Haldane1988, Tanaka2005, Senthil2006}.

A critical system with a lower-dimensional boundary shows rich surface critical behavior, which is controlled by the interactions near the boundary. The surface criticality of the classical phase transitions has been well-documented in the literature \cite{Binder1983phase, Diehl1986phase}, which can be classified based on whether the boundary has a long-range order at the bulk critical point. In the ordinary class, the surface remains disordered throughout the disordered phase in the bulk, and the surface critical behavior is fully induced by the divergent correlation length in the bulk. If the surface interaction is so strong that it develops a long-range order at a higher temperature than the bulk, the surface exhibits additional critical behavior at the bulk transition, which is classified as the extraordinary transition. Fine-tuning the surface interaction can lead to a special transition, where the surface and the bulk ordering transitions merge at the same temperature.

The interest in the surface critical behavior particularly as QCPs has revived recently in order to unveil the physical properties of the gapless or degenerate surface modes of topological states as the bulk undergoes quantum phase transitions (QPTs). It is shown that in certain models of topological superconductors the Majorana edge modes are robust against the critical quantum fluctuations in the bulk \cite{Cheng2011, Fidkowski2011, Sau2011, Grover2012a}. In the two-dimensional (2D) Affleck-Kennedy-Lieb-Tasaki model, the gapless edge state changes the universality class of the surface critical behavior of the (2+1)D O(3) Wilson-Fisher QCP \cite{Zhang2017}. This new surface universality class is also realized in similar models with dangling spin chains on the edge \cite{Ding2018, Weber2018, Weber2019a, Zhu2021} and has triggered further theoretical studies \cite{Ding2021, ParisenToldin2021, Hu2021, ParisenToldin2022, Jian2021, Metlitski2022, Weber2021, Zhu2021b, Yu2022}.

The notion of symmetry-protected topological (SPT) states has been generalized into the higher-order (HO) topological states with lower-dimensional edge modes \cite{Benalcazar2017a, Benalcazar2017, Song2017a, Langbehn2017, Schindler2018, Trifunovic2019, Wang2018d, You2018a, Dubinkin2019}. A simple example is provided by the spin-$1/2$ antiferromagnetic Heisenberg model on a plaquette-modulated square lattice, which is shown in Fig. \ref{fig:model}. For $J_{2}/J_{1}\ll 1$, all spins in the bulk form plaquette singlets thus the bulk is in a gapped disordered phase. With the open boundary condition, the dimerized edge is also gapped, but corner A in Fig. \ref{fig:model} hosts a dangling spin-$1/2$ mode, whose degeneracy is protected by spin rotation, lattice $C_{4}$ rotation and $R_{xy}$ reflection symmetries \cite{You2018a, Dubinkin2019, Peng2021}. Therefore, this plaquette Heisenberg model realizes a second-order SPT state with a spin-$1/2$ corner mode. The HOSPT phase and the lower-dimensional boundary mode can be captured by a topological term in the effective field theory \cite{You2018a, Peng2021}. Tuning the interaction $J_{2}/J_{1}$ induces a QPT from the disordered phase to an antiferromagnetic order in the $J_{2}/J_{1}\simeq 1$ regime, which belongs to the (2+1)D O(3) universality class \cite{Wenzel2008, Wenzel2009, Ran2019, Xu2019a}. The direct topological QPT between different disordered phases was also studied recently \cite{Peng2021}.

In this work, we study the critical behavior of the corner state at the bulk QCP between the disordered and the antiferromagnetic ordered phases, and focus on the difference of the corner states with or without a dangling spin. With quantum Monte Carlo simulations on lattices with open boundary condition, we find such two corner states exhibit distinct critical behavior with different critical exponents. In particular, we find that the degenerate spin-$1/2$ mode at the dangling corner persists at the QCP despite its coupling to the critical spin fluctuations in the bulk, which is reminiscent of the extraordinary class of the classical surface critical behavior.

The paper is organized as follows: In Sec. \ref{sec:model}, we introduce the model and the spin correlation functions used to characterize the corner critical behavior. In Sec. \ref{sec:results}, the correlation functions of the dangling and the nondangling corner spins are presented, from which the critical exponents are extracted. A summary is given in Sec. \ref{sec:con}.

\section{Model and methods}
\label{sec:model}

\begin{figure}[tb]
\includegraphics[width=\columnwidth]{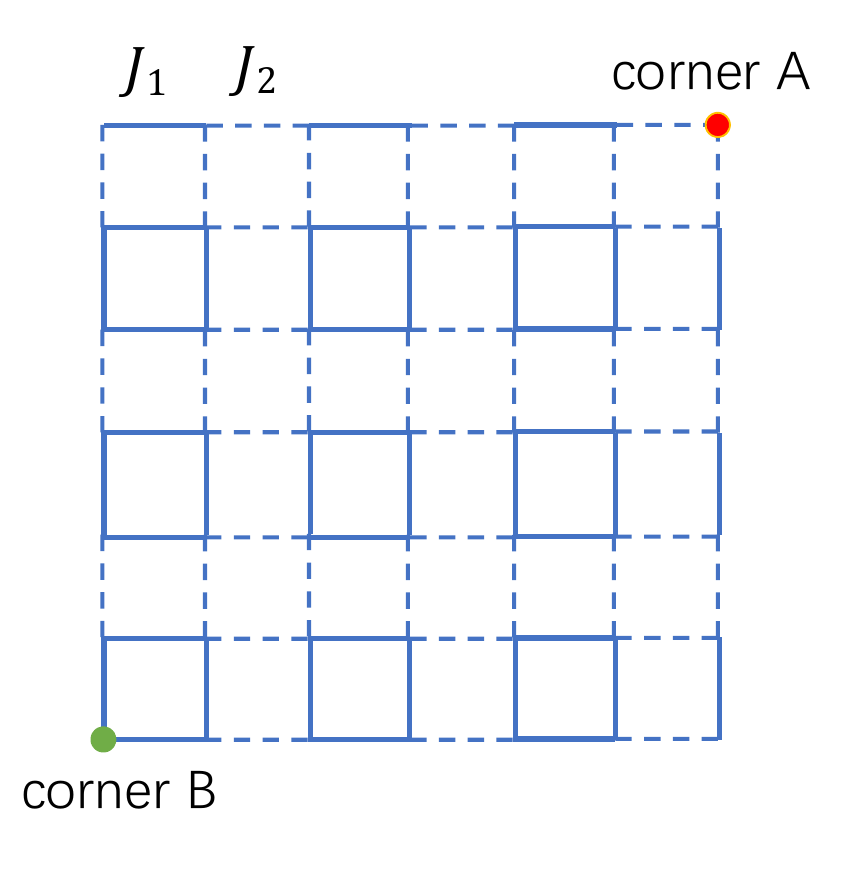}
\caption{Structures of the plaquette square lattice. Each unit cell (a plaquette) is composed of four sites. The intra-unit-cell interaction $J_{1}$ and the inter-unit-cell interaction $J_{2}$ are represented by solid and dashed bonds, respectively. There is a dangling spin corner A with a degenerate spin-$1/2$ mode in the $J_{2}/J_{1}\ll 1$ regime, while corner B is called the nondangling corner.}
\label{fig:model}
\end{figure}

We study the $S=1/2$ antiferromagnetic Heisenberg model on the plaquette square lattice shown in Fig.\ref{fig:model}. Each unit cell (a plaquette) is composed of four sites. The Hamiltonian is given by
\begin{equation}
H=J_{1}\sum_{\langle i,j\rangle}\mathbf{S}_{i}\cdot \mathbf{S}_{j}+J_{2}\sum_{\langle i,j\rangle'}\mathbf{S}_{i}\cdot \mathbf{S}_{j},
\label{Hamiltonian}
\end{equation}
where $J_{1}$ and $J_{2}$ are the nearest-neighbor antiferromagnetic Heisenberg interactions of intra- and inter-unit-cell bonds, respectively. In the $J_{2}/J_{1}\ll 1$ limit, each plaquette forms a spin singlet state at the ground state, thus the system is in a gapped disordered phase, which is called the plaquette valence bond solid (PVBS) phase. For $J_{2}/J_{1}\simeq 1$, the system has a long-range N\'eel order. There is a quantum phase transition from the PVBS phase to the N\'eel order at $J_{2}/J_{1}=0.548524$, which belongs to the 3D O(3) universality class \cite{Wenzel2008, Wenzel2009, Ran2019, Xu2019a}. In this work, we set $J_{1}=1$ as the unit of energy.

It is realized recently that this plaquette Heisenberg model is an example of HOSPT phase \cite{You2018a, Dubinkin2019, Peng2021}. In the disordered phase, as shown in Fig. \ref{fig:model}, the lattice with open boundary has a dangling spin $1/2$ at corner A, because all strong bonds connecting to the corner site are cut. This dangling spin $1/2$ forms a degenerate corner mode, which is protected by the spin rotation symmetry together with the crystal symmetry: the plaquette-centered $C_{4}$ rotation and the diagonal reflection $R_{xy}$ symmetry. On the other hand, the spin on corner B forms a singlet with its neighboring sites, and is not degenerate.

In this work, we use the stochastic series expansion (SSE) quantum Monte Carlo (QMC) simulations \cite{Sandvik1992a, Sandvik1999, Dorneich2001, Syljuasen2002} to study the corner critical behavior of the plaquette Heisenberg model with open boundary. The lattice size is $L\times L$ with $33\leq L\leq 81$. The inverse temperature $\beta=L$. All simulations are performed at the bulk quantum critical point (QCP) $J_{2}/J_{1}=0.548524$.

The following spin correlation functions and the corresponding anomalous dimensions are introduced to characterize the critical behavior of the corner spin at the QCP: the equal-time correlation of the corner spin (with subscript $c$) at $(0,0)$ and a bulk spin (with subscript $b$) at $(r,r)$,
\begin{equation}
C_{cb}(r)=\langle S_{r}^{z}S_{0}^{z}\rangle \propto r^{-(d+z-2+\eta_{cb})},
\end{equation}
and the equal-time correlation of the corner spin at $(0,0)$ and a spin on the edge (with subscript $s$) at $(r,0)$,
\begin{equation}
C_{cs}(r)=(-1)^{r}\langle S_{r}^{z}S_{0}^{z}\rangle \propto r^{-(d+z-2+\eta_{cs})},
\end{equation}
and the imaginary-time correlation of the spin at the corner,
\begin{equation}
C_{cc}(\tau)=\langle S^{z}(\tau)S^{z}(0)\rangle \propto \tau^{-(d+z-2+\eta_{cc})/z}.
\label{eq:fss-tau}
\end{equation}
Here, $d=2$ is the spatial dimension, and $z=1$ is the dynamical exponent of the QCP.

These anomalous dimensions can be derived from the scaling dimensions of the spin operators in the bulk, on the edge and at the corner, which are denoted by $\Delta_{b}$, $\Delta_{s}$ and $\Delta_{c}$, respectively:
\begin{align}
d+z-2+\eta_{cb} &= \Delta_{c}+\Delta_{b}, \\
d+z-2+\eta_{cs} &= \Delta_{c}+\Delta_{s}, \\
d+z-2+\eta_{cc} &= 2\Delta_{c}.
\end{align}
Moreover, the anomalous dimensions in the bulk $\eta$ and on the edge $\eta_{\parallel}$ are defined by
\begin{align}
d+z-2+\eta &= 2\Delta_{b}, \\
d+z-2+\eta_{\parallel} &= 2\Delta_{s},
\end{align}
thus we have the following scaling relations,
\begin{align}
2\eta_{cb} &= \eta+\eta_{cc}, \label{eq:rel1} \\
2\eta_{cs} &= \eta_{\parallel}+\eta_{cc}. \label{eq:rel2}
\end{align}
These relations will serve as consistency check to the following numerical results. For the 3D O(3) universality class $\eta= 0.0386(12)$ \cite{Kos2016, Poland2019}. The edge is always gapped in the paramagnetic phase, thus the surface critical behavior belongs to the ordinary class with $\eta_{\parallel}=1.338$ \cite{Diehl1994, Diehl1998, Gliozzi2015}.

The spatial correlations $C_{cb}(r)$ and $C_{cs}(r)$ show even-odd effect as a function of the distance $r$ due to the plaquette modulation. Therefore, we fit the correlation functions of a given lattice size $L$ for even and odd distances $r$ separately to extract the anomalous dimensions $\eta_{i}^{e}(L)$ and $\eta_{i}^{o}(L)$ ($i=cb$ or $cs$), and extrapolate to the thermodynamic limit with the following scaling form \cite{Luck1985},
\begin{equation}
\eta_{i}^{e/o}(L) = \eta_{i}^{e/o}+cL^{-\omega}.
\label{eq:correction}
\end{equation}
The extrapolated exponents $\eta_{i}^{e}$ and $\eta_{i}^{o}$ are expected to be consistent with each other. In practice, their slight difference provides an estimate of the systematic error in the finite-size scaling analysis.

\section{Numerical results}
\label{sec:results}

\subsection{Nondangling corner}

\begin{figure}[tb]
\includegraphics[width=\columnwidth]{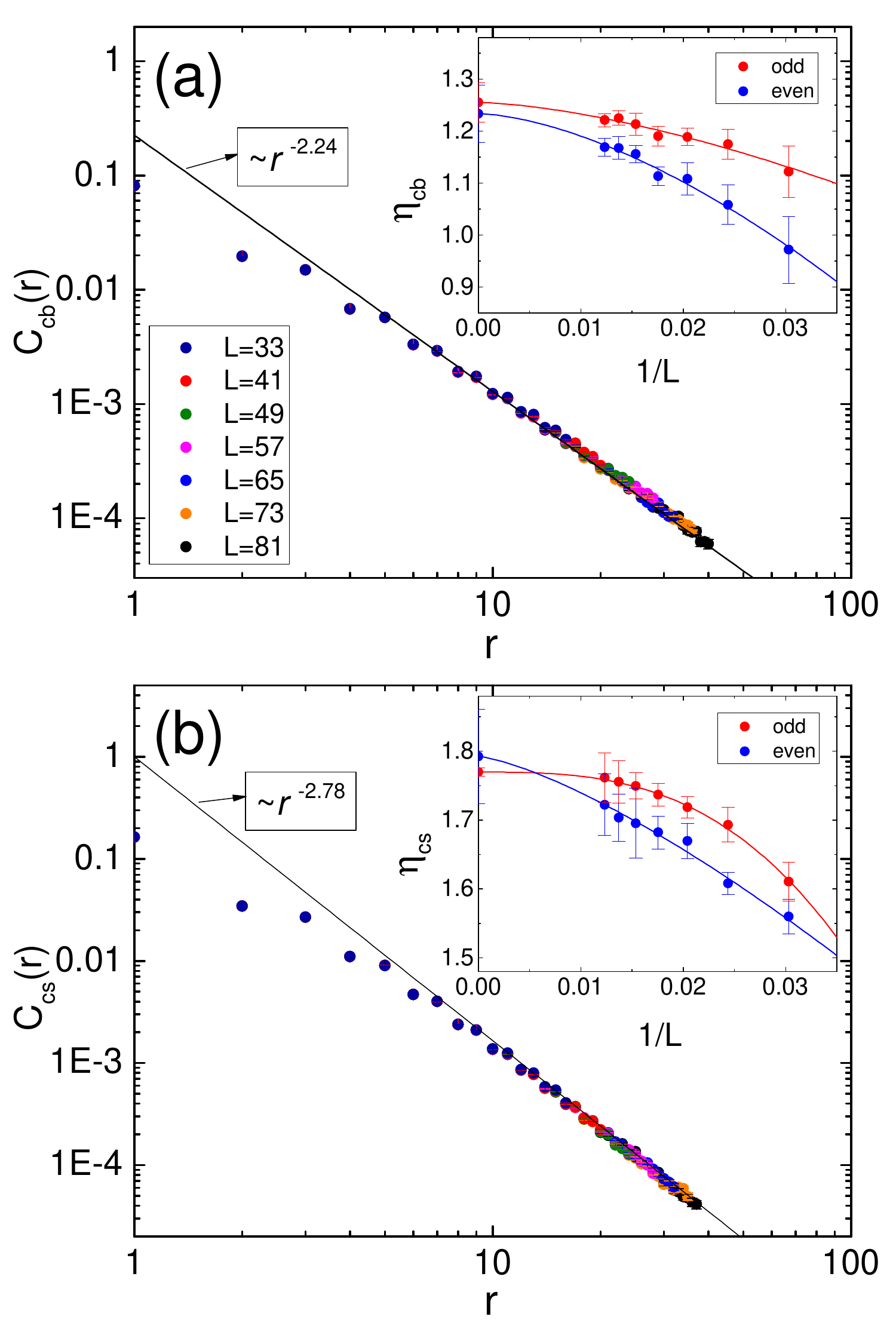}
\caption{Spatial correlation functions (a) $C_{cb}(r)$ and (b) $C_{cs}(r)$ of the nondangling corner spin. The black lines are the reference for the power-law decay with the extrapolated critical exponents, $\eta_{cb}=1.24(6)$ and $\eta_{cs}=1.78(5)$. Insets: the finite-size scaling of the critical exponents $\eta_{cb}$ and $\eta_{cs}$ with Eq. (\ref{eq:correction}). The correction-to-scaling exponents $\omega_{cb}^{e}=1.6(8)$, $\omega_{cb}^{o}=1.6(9)$, $\omega_{cs}^{e}=1.3(7)$, and $\omega_{cs}^{o}=2.8(6)$.
}
\label{fig:B-cbcs}
\end{figure}

\begin{ruledtabular}
\begin{table}[tb]
\caption{Anomalous dimensions of the corner critical behavior for both the nondangling and the dangling corner spins.}
\label{tab:exponents}
\begin{tabular}{c|cccc}
			& $\eta_{cb}$	& $\eta_{cs}$	& $\eta_{cc}$	& $\Delta_{c}$	\\
\hline
Nondangling	& 1.24(6)		& 1.78(5)		& 2.34(3)		& 1.67(2)		\\
Dangling	& 0.33(3)		& 1.06(3)		& 0.682(11)		& 0.84(1)				
\end{tabular}
\end{table}
\end{ruledtabular}

\begin{figure}[tb]
\includegraphics[width=\columnwidth]{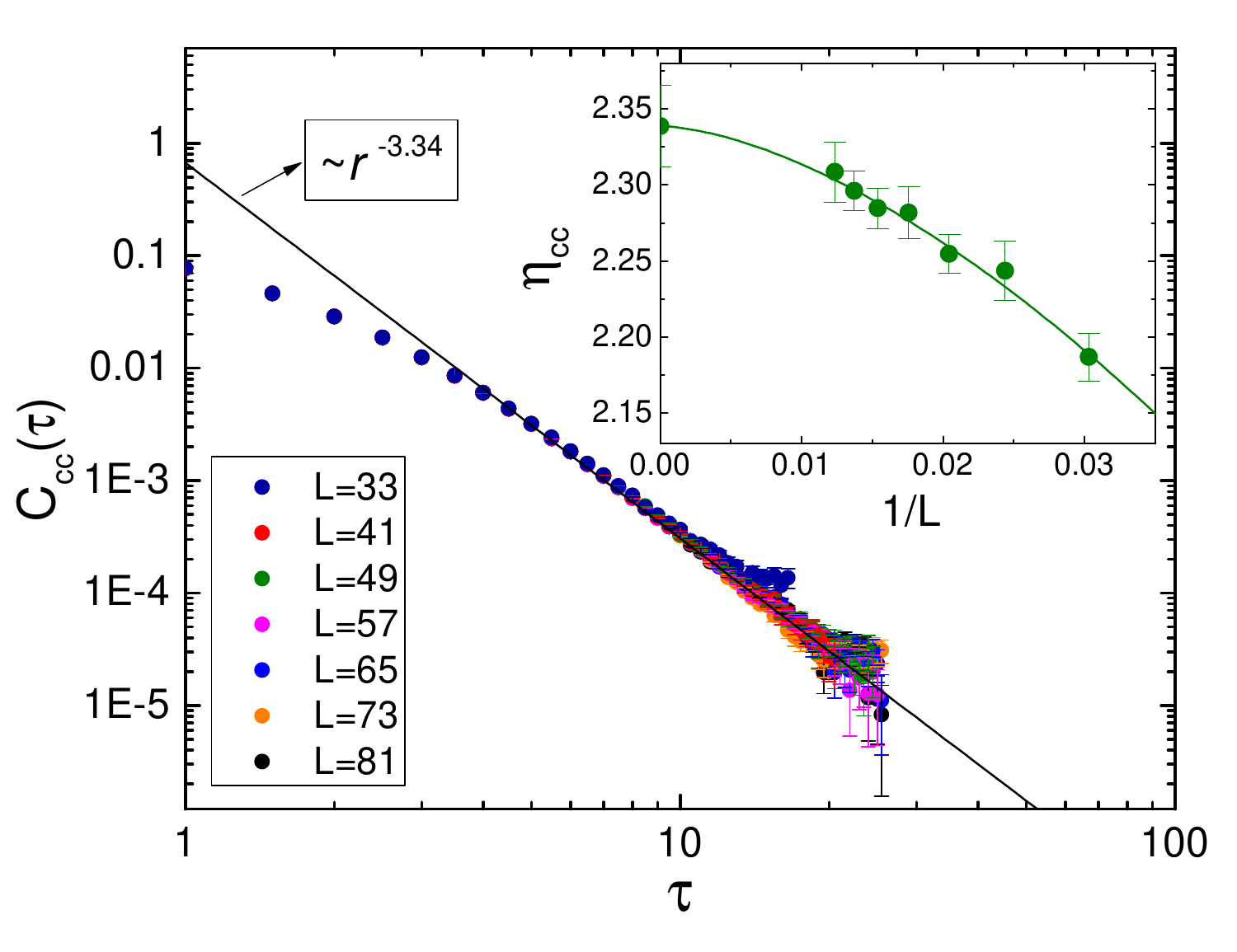}
\caption{Imaginary-time correlation function $C_{cc}(\tau)$ of the nondangling corner spin. The black line is the reference for the power-law decay with the extrapolated critical exponent, $\eta_{cc}=2.34(3)$. Inset: the finite-size scaling of the critical exponent $\eta_{cc}$ with Eq. (\ref{eq:correction}). The correction-to-scaling exponent $\omega_{cc}=1.6(6)$.
}
\label{fig:B-cc}
\end{figure}

We first study the critical behavior of the nondangling corner (corner B in Fig. \ref{fig:model}), which is dubbed the ordinary transition of the corner criticality, because the corner spin is gapped in the disordered phase. The correlation functions $C_{cb}(r)$ and $C_{cs}(r)$ are shown in Fig. \ref{fig:B-cbcs}. Using the finite-size scaling analysis introduced in Sec. \ref{sec:model}, we find the following anomalous dimensions for $C_{cb}(r)$, $\eta_{cb}^{e}=1.23(6)$, and $\eta_{cb}^{o}=1.26(4)$, which are consistent with each other within error bars. The average $\eta_{cb}=\frac{1}{2}(\eta_{cb}^{e}+\eta_{cb}^{o})$ is taken as an estimate of the exponent. An estimate of the error bar $\epsilon_{cb}$ is given by $\epsilon_{cb}=\frac{1}{2}(\epsilon_{cb}^{e}+\epsilon_{cb}^{o}+|\eta_{cb}^{e}-\eta_{cb}^{o}|)$. The final estimate is
\begin{equation}
\eta_{cb}^{\mr{(ord)}}=1.24(6).
\end{equation}
With similar analysis, we find for $C_{cs}(r)$, $\eta_{cs}^{e}=1.79(7)$, and $\eta_{cs}^{o}=1.77(1)$, and our final estimate is
\begin{equation}
\eta_{cs}^{\mr{(ord)}}=1.78(5).
\end{equation}
The imaginary time correlation function of the corner spin $C_{cc}(\tau)$ is shown in Fig. \ref{fig:B-cc}. Fitting to Eq. (\ref{eq:fss-tau}) and then Eq. (\ref{eq:correction}), we find the anomalous dimension
\begin{equation}
\eta_{cc}^{\mr{(ord)}}=2.34(3).
\end{equation}
These anomalous dimensions are listed in Table \ref{tab:exponents}. The scaling relations in Eq. (\ref{eq:rel1}) and (\ref{eq:rel2}) are respected within error bars. Therefore, our final estimate of the scaling dimension of the corner spin operator at the ordinary transition is
\begin{equation}
\Delta_{c}^{\mr{(ord)}}=1.67(2).
\end{equation}

\subsection{Dangling corner}

\begin{figure}[tb]
\includegraphics[width=\columnwidth]{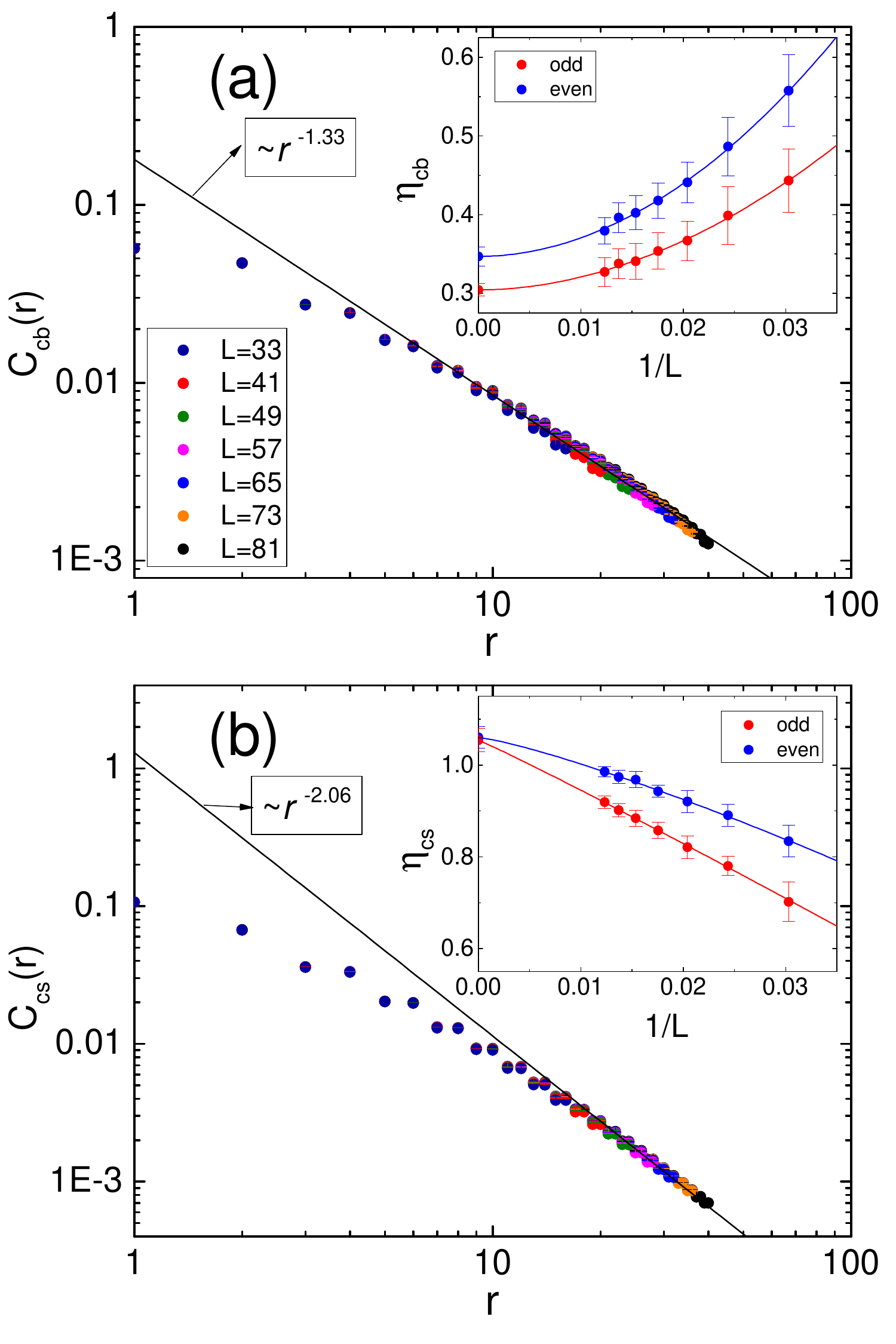}
\caption{Spatial correlation functions (a) $C_{cb}(r)$ and (b) $C_{cs}(r)$ of the dangling corner spin. The black lines are the reference for the power-law decay with the extrapolated critical exponents, $\eta_{cb}=0.33(3)$ and $\eta_{cs}=1.06(3)$. Insets: the finite-size scaling of the critical exponents $\eta_{cb}$ and $\eta_{cs}$ with Eq. (\ref{eq:correction}). The correction-to-scaling exponents are given by $\omega_{cb}^{e}=2.0(3)$, $\omega_{cb}^{o}=1.9(3)$, $\omega_{cs}^{e}=1.2(3)$, and $\omega_{cs}^{o}=1.0(2)$.
}
\label{fig:A-cbcs}
\end{figure}

\begin{figure}[tb]
\includegraphics[width=\columnwidth]{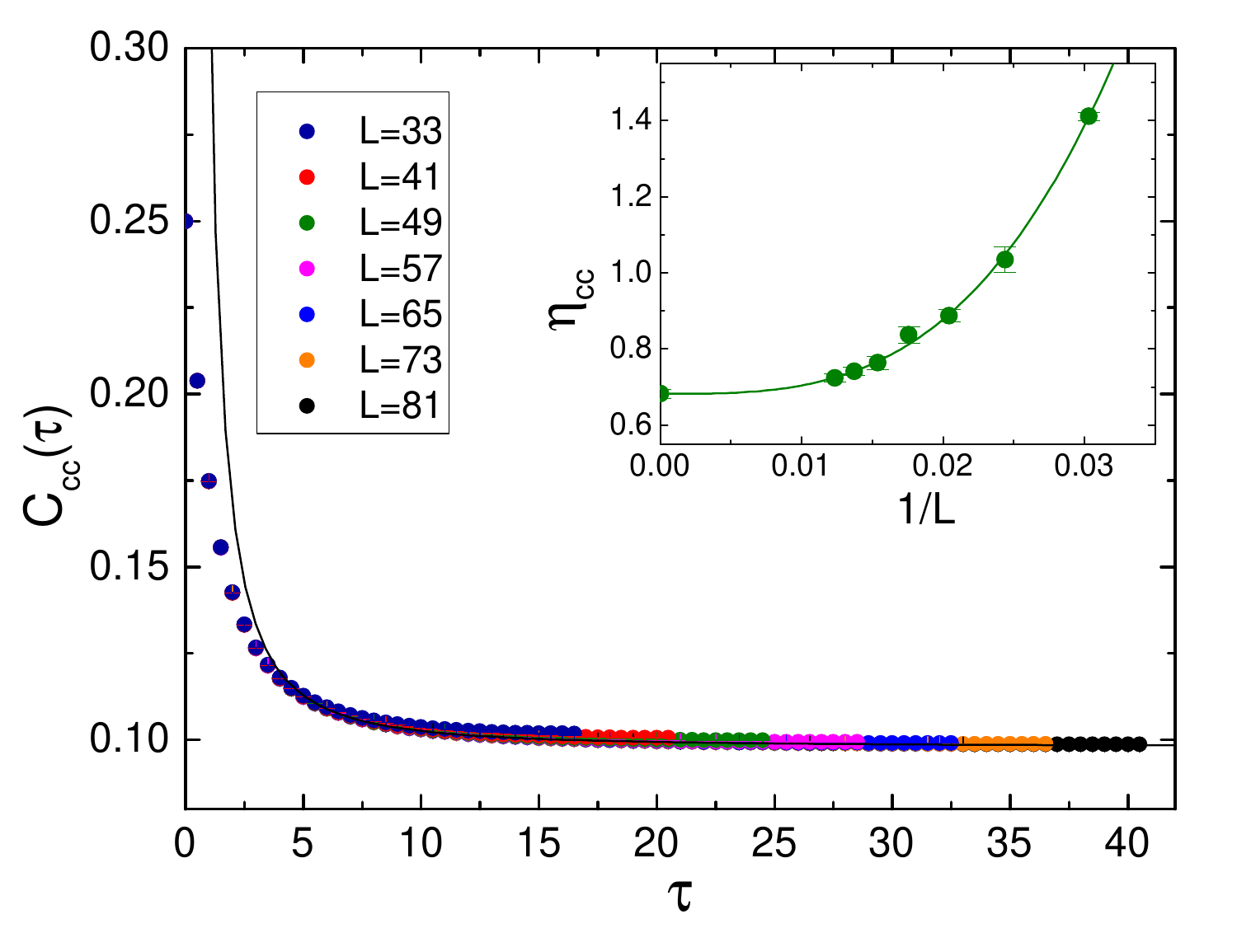}
\caption{Imaginary-time correlation function $C_{cc}(\tau)$ of the dangling corner spin, which does not decay to zero, indicating a persistent degenerate spin mode. Inset: the finite-size scaling of the critical exponent $\eta_{cc}$ with Eq. (\ref{eq:correction}). The correction-to-scaling exponent $\omega_{cc}=3.2(2)$.
}
\label{fig:A-cc}
\end{figure}

The dangling corner has a degenerate spin-$1/2$ mode in the bulk disordered phase, which spontaneously breaks the spin rotational symmetry at the corner. We thus expect that it leads to corner critical behavior distinct from the ordinary class. The spatial correlation functions $C_{cb}(r)$ and $C_{cs}(r)$, and the imaginary-time correlation $C_{cc}(\tau)$ are shown in Figs. \ref{fig:A-cbcs} and \ref{fig:A-cc}, respectively.

We first note that $C_{cc}(\tau)$ does not decay to zero as $\tau$ grows (see Fig. \ref{fig:A-cc}), indicating a persistent spin mode at the corner despite its coupling to the critical bulk state. We stress that the persistent large-$\tau$ spin correlation does not depend on whether the total number of lattice sites is even or odd as long as the corner site is coupled to the bulk with weak bonds. We thus include a constant term in the fitting,
\begin{equation}
C_{cc}(\tau)=m_{c}^{2}+b\tau^{-1-\eta_{cc}}.
\label{eq:constant}
\end{equation}
The anomalous dimension $\eta_{cc}$ obtained from different lattice sizes is then extrapolated to the thermodynamic limit with Eq. (\ref{eq:correction}). The result is
\begin{equation}
\eta_{cc}^{\mr{(ext)}}=0.682(11).
\end{equation}
The constant term in Eq. (\ref{eq:constant}) indicates a free local moment at the corner, which can be extrapolated with
\begin{equation}
m_{c}^{2}(L)=m_{c}^{2}+bL^{-\omega_{c}},
\end{equation}
and we find $m_{c}^{2}=0.0974(2)$, and the correction-to-scaling exponent $\omega_{c}=1.42(7)$. Therefore, we may dub the corner critical behavior with a persistent corner spin mode the extraordinary class. The corner spin coupled to the critical state in the bulk may be treated as a Bose-Kondo problem by analogy with the Kondo problem, in which a local moment couples to gapless electrons in the metal. In the Kondo problem in metals, the local moment is screened in the low-energy limit; however, in our case, the local moment, i.e., the corner spin remains free despite its coupling to the critical bulk state.

The spatial correlations $C_{cb}(r)$ and $C_{cs}(r)$ show simple power-law scaling form. With the finite-size scaling analysis sketched in Sec. \ref{sec:model}, we find $\eta_{cb}^{e}=0.347(11)$, $\eta_{cb}^{o}=0.304(9)$, and $\eta_{cs}^{e}=1.06(3)$, $\eta_{cs}^{o}=1.05(3)$. Our final estimate of these exponents are
\begin{equation}
\eta_{cb}^{\mr{(ext)}}=0.33(3),\quad \eta_{cs}^{\mr{(ext)}}=1.06(3).
\end{equation}

The scaling relations in Eqs. (\ref{eq:rel1}) and (\ref{eq:rel2}) are respected within error bars, thus our estimate of the scaling dimension of the corner spin operator is
\begin{equation}
\Delta_{c}^{\mr{(ext)}}=0.841(6).
\end{equation}

\begin{figure}[tb]
\centering
\includegraphics[width=\columnwidth]{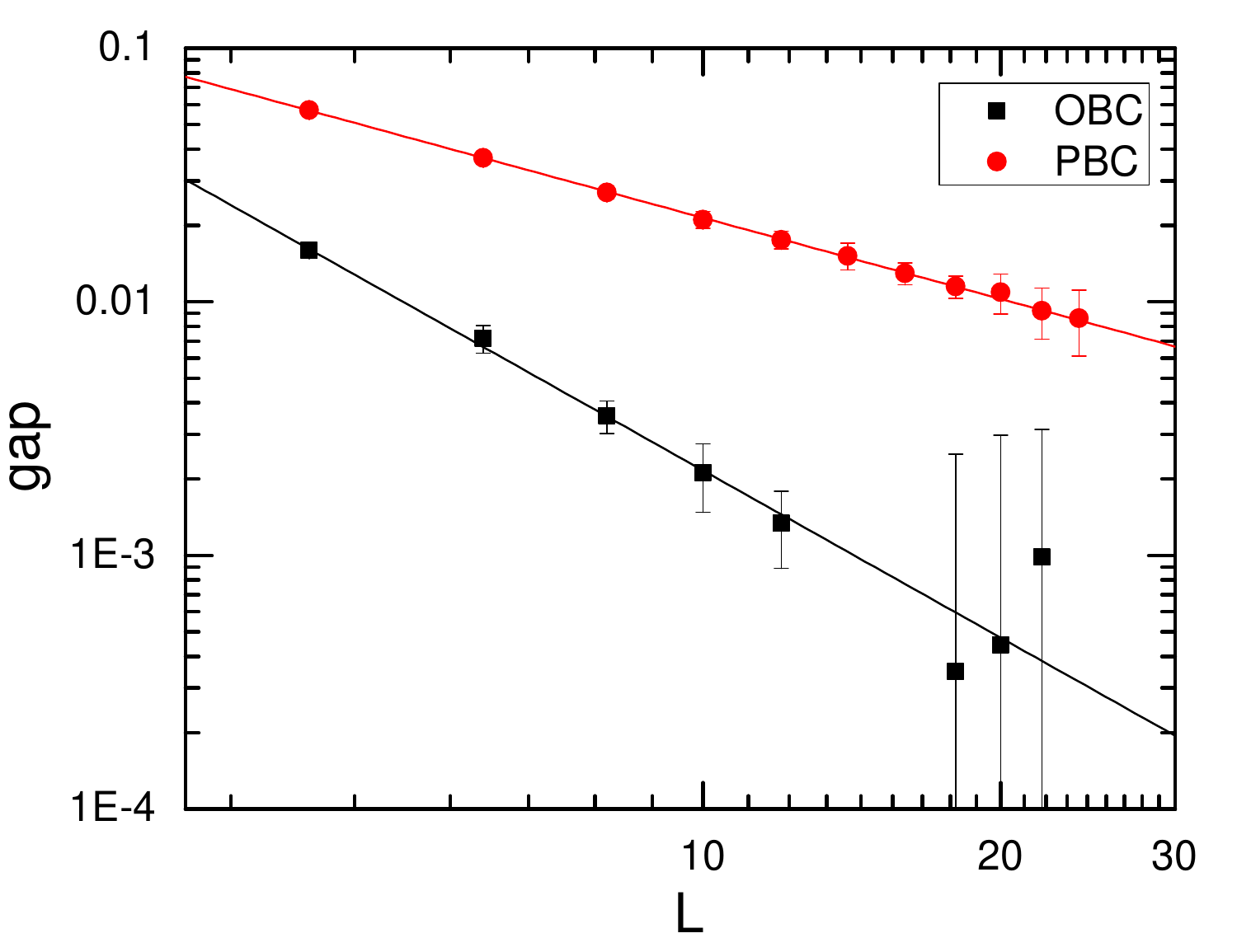}
\caption{The spin excitation gaps of the plaquette Heisenberg model on finite-size lattices. The lines are power-law fittings. For the periodic boundary condition (PBC, red circles), the gap $\Delta_{b}(L)\propto L^{-z}$ with $z=1.064(7)$, which is consistent with the dynamic exponent $z=1$ of the (2+1)D O(3) QCP. For the open boundary condition with dangling corners (OBC, black squares), the gap $\Delta_{c}(L)\propto L^{-\omega}$ with $\omega=2.19(7)$. }
\label{fig:gap}
\end{figure}

The spin excitation gaps on finite-size lattices are extracted by calculating the ground state energies in both the spin singlet and triplet sectors with projective QMC algorithm. The results are shown in Fig. \ref{fig:gap}. The finite-size gap of the bulk excitation is obtained with the periodic boundary condition, which scales as $L^{-z}$, in which $z=1$ is the dynamic exponent of the QCP. The gap in the presence of dangling corner spins with open boundary condition is much smaller: it decays as $L^{-\omega}$ with $\omega=2.19(7)$. This suggests that there are even lower energy states associated with the dangling corners besides the gapless excitations in the bulk. In the thermodynamic limit $L\rightarrow \infty$, these corner modes become truly free spin-$1/2$ states and effectively decouple from the bulk critical states, which is consistent with the irrelevance of the bulk-corner coupling, as we shall discuss with the dimensional and RG analysis below.

The dangling corner at the bulk QCP can be modeled by a free spin $1/2$ coupled to the bulk state with an ordinary corner criticality. The action is given by
\begin{equation}
S=S_{0}[\hat{n}]+S_{b}[\vec{\phi}]+S_{c}[\hat{n},\vec{\phi}],
\label{eq:s}
\end{equation}
in which $S_{0}[\hat{n}]$ is the $(0+1)$-D Wess-Zumino-Witten action of a free spin $1/2$, which can obtained from the spin coherent state path integral \cite{Altland2010condensed},
\begin{equation}
S_{0}[\hat{n}]=iS\int\mr{d}\tau\int_{0}^{1}\mr{d}u \hat{n}\cdot (\pd_{\tau}\hat{n}\times \pd_{u}\hat{n})
\end{equation}
with $S=1/2$, in which the unit vector $\hat{n}(\tau)$ represents the direction of the corner spin, which is lifted to a continuous field $\hat{n}(\tau,u)$ over the imaginary time $\tau$ and an auxiliary parameter $u\in [0,1]$. The lifting is an arbitrary continuous mapping satisfying
\begin{equation}
\hat{n}(\tau,u=0)=(0,0,1),\quad \hat{n}(\tau,u=1)=\hat{n}(\tau).
\end{equation}
The $S_{b}[\vec{\phi}]$ term captures the quasi-long-range spin correlation of the bulk order parameter near the corner $\vec{\phi}(\tau)$,
\begin{equation}
S_{b}[\vec{\phi}]=\frac{1}{2}\int\mr{d}\tau\mr{d}\tau' K^{-1}(\tau,\tau')\vec{\phi}(\tau)\cdot\vec{\phi}(\tau'),
\end{equation}
in which
\begin{equation}
K(\tau,\tau')\propto|\tau-\tau'|^{-2\Delta_{c}^{(\mr{ord})}}
\label{eq:K}
\end{equation}
is the temporal spin correlation function of the ordinary corner criticality, and $K^{-1}(\tau,\tau')$ is the inverse matrix of $K(\tau,\tau')$. The coupling of the dangling spin to the bulk is given by
\begin{equation}
S_{c}[\hat{n},\vec{\phi}]=-\lambda\int\mr{d}\tau \hat{n}(\tau)\cdot\vec{\phi}(\tau).
\end{equation}
The unit vector $\hat{n}(\tau)$ is dimensionless, $[\hat{n}]=0$, while the bulk order parameter at an ordinary corner criticality has the scaling dimension $[\vec{\phi}]=1.67(2)$, thus a simple dimension counting suggests that the coupling term $S_{c}$ is irrelevant and the dangling corner spin remains free despite its coupling to the bulk critical state.

In a recent work \cite{Nahum2022}, a quantum spin with a long-range temporal interaction was studied with the RG analysis. Its action
\begin{equation}
S[\hat{n}]= \frac{1}{2g}\int\mr{d}\tau\mr{d}\tau'K(\tau-\tau')\big(\hat{n}(\tau)-\hat{n}(\tau')\big)^{2}+S_{0}[\hat{n}]
\end{equation}
can be obtained from Eq. (\ref{eq:s}) by integrating out $\vec{\phi}(\tau)$, with $g^{-1}=\lambda^{2}$. The RG equation is governed by the exponent in Eq. (\ref{eq:K}). In particular, for $\Delta_{c}^{(\mr{ord})}>1$, the coupling $\lambda$ flows to zero, thus the dangling spin is free in the low-energy limit, which is consistent with our numerical results.

\section{Conclusion}
\label{sec:con}

To summarize, we have studied the corner critical behavior at the QCP of a plaquette Heisenberg model with extensive quantum Monte Carlo simulations. We find that the two corners with or without a dangling spin show different critical exponents, thus belong to different universality classes. In particular, the spin-$1/2$ mode at the dangling corner, which is a hallmark of the HOSPT state in the disordered phase, remains free with long-range temporal correlation at the bulk QCP despite its coupling to the critical spin fluctuations.

\begin{acknowledgements}
We thank Xue-Feng Zhang and Dong-Xu Liu for helpful discussion. This work is supported by the National Key R\&D Program of China (2018YFA0305800), the National Natural Science Foundation of China (11804337, 12047564 and 12174387), the Strategic Priority Research Program of CAS (XDB28000000), the Fundamental Research Funds for the Central Universities under Grant No. 2020CDJQY-Z003 and the CAS Youth Innovation Promotion Association.
\end{acknowledgements}

\bibliography{../../../BibTex/library,../../../BibTex/books}
\end{document}